\documentclass{elsart}
\journal{Physics Letters B}
\usepackage{graphics}
\usepackage{graphicx}
\usepackage{epsfig}
\usepackage{amssymb}
\usepackage{amsmath}
\usepackage{textcomp}
\usepackage{dcolumn}
\usepackage{array}
\usepackage{varioref}
\usepackage{afterpage}

\begin{document}

\begin{frontmatter}

\begin{flushleft}
\hspace*{9cm}BELLE Preprint 2007-28 \\
\hspace*{9cm}KEK Preprint 2007-17 \\
\end{flushleft}

\title{ Study of $\tau^-\to K_S \pi^-\nu_{\tau}$ decay at Belle }

\collab{Belle Collaboration}
  \author[BINP]{D.~Epifanov}, % BINP 
  \author[KEK]{I.~Adachi}, % KEK
  \author[Tokyo]{H.~Aihara}, % Tokyo
  \author[BINP]{K.~Arinstein}, % BINP
  \author[BINP]{V.~Aulchenko}, % BINP
  \author[Lausanne,ITEP]{T.~Aushev}, % ITEP
  \author[Sydney]{A.~M.~Bakich}, % Sydney
  \author[ITEP]{V.~Balagura}, % ITEP
  \author[Melbourne]{E.~Barberio}, % Melbourne
  \author[BINP]{I.~Bedny}, % BINP
% ...
% Ljubljana

\address[BINP]{Budker Institute of Nuclear Physics, Novosibirsk, Russia}
% ...
\address[KEK]{High Energy Accelerator Research Organization (KEK), Tsukuba, Japan}
% ...
\address[ITEP]{Institute for Theoretical and Experimental Physics, Moscow, Russia}
% ...
\address[Lausanne]{Swiss Federal Institute of Technology of Lausanne, EPFL, Lausanne, Switzerland}
% ...
\address[Melbourne]{University of Melbourne, Victoria, Australia}
% ...
\address[Sydney]{University of Sydney, Sydney, NSW, Australia}
% ...
\address[Tokyo]{Department of Physics, University of Tokyo, Tokyo, Japan}
% ...

\begin{abstract}
We present a study of the decay $\tau^-\to K_S \pi^-\nu_{\tau}$ 
using a $351\ {\rm fb^{-1}}$ data sample collected with the Belle detector.
The analysis is based on $53,110$ lepton-tagged signal events. 
The measured branching fraction
$\mathcal{B}(\tau^-\to K_S\pi^-\nu_{\tau})=
(0.404\pm 0.002({\rm stat.})\pm 0.013({\rm syst.}))\%$
is consistent with the world average value and has better accuracy.
An analysis of the $K_S\pi^-$ invariant mass spectrum 
reveals contributions from the $K^*(892)^-$ as well
as other states. For the first time the $K^*(892)^-$ mass and width have been 
measured in $\tau$ decay:
$M(K^*(892)^-)=(895.47\pm 0.20({\rm stat.}) \pm 0.44({\rm syst.}) \pm 
0.59({\rm mod.}))\ {\rm MeV/}c^2$, 
$\Gamma(K^*(892)^-)=(46.2\pm 0.6({\rm stat.})\pm 1.0({\rm syst.})\pm 
0.7({\rm mod.}))\ {\rm MeV}$. The $K^*(892)^-$ mass is significantly
different from the current world average value.   
\end{abstract}


\end{frontmatter}

\section{Introduction}

$\tau$ lepton hadronic decays provide a laboratory for the
study of low energy hadronic currents under very clean conditions.
In these decays, the hadronic system is produced from the QCD vacuum 
via the charged weak current mediated by a $W^{\pm}$ boson. 
The $\tau$ decay amplitude can thus be factorized into a purely leptonic 
part including the $\tau$ and $\nu_{\tau}$ and a hadronic spectral function.
Strangeness changing $\tau$ decays are suppressed by a factor of
$\simeq 20$ relative to Cabibbo-allowed modes. High-statistics
measurements at $B$ factories provide excellent opportunities for studying
the structure of the strange hadronic spectral functions in specific decay 
modes \cite{stone,sttwo,stthr}, the parameters of the intermediate states 
and the total strange hadronic spectral function \cite{spe}.

The decay $\tau^- \to \bar{K^0} \pi^- \nu_{\tau}$ (unless specified
otherwise, charge conjugate decays are implied throughout the paper)
has the largest branching
fraction of all Cabibbo-suppressed decays of the $\tau$ lepton.
Early studies of this decay established that the main 
contribution to the $K\pi$ invariant mass spectrum is from the 
$K^*(892)$ meson~\cite{dorfan,argus,cltwo}. 
Although scalar or tensor contributions are expected in theoretical
models~\cite{fione,gone} and not excluded experimentally~\cite{alone,clone}, 
the low statistics of previous investigations did not allow for a detailed study.

Here we report a precise measurement of the branching fraction for the decay
$\tau^- \to K_S \pi^- \nu_{\tau}$ as well as a study of its final state dynamics. 
This analysis  is based on a $351~{\rm fb}^{-1}$ data sample that
contains 313 $\times 10^6\ \tau^+\tau^-$ pairs, 
collected  with the Belle detector at the KEKB energy-asymmetric 
$e^+e^-$ (3.5 on 8~GeV) collider~\cite{kekb}
operating at the $\Upsilon(4S)$ resonance.

\section{The Belle detector}

The Belle detector is a large-solid-angle magnetic spectrometer that
consists of a silicon vertex detector (SVD),
a 50-layer central drift chamber (CDC), an array of
aerogel threshold Cherenkov counters (ACC), 
a barrel-like arrangement of time-of-flight
scintillation counters (TOF), and an electromagnetic calorimeter (ECL)
comprised of CsI(Tl) crystals located inside 
a superconducting solenoid coil that provides a 1.5~T
magnetic field.  An iron flux-return located outside 
the coil is instrumented to detect $K_L^0$ mesons and to identify
muons (KLM).  
Two inner detector configurations are used in this analysis. A beampipe 
with a radius of 2.0 cm and a 3-layer silicon vertex detector 
are used for the first sample
of 124 $\times 10^6\ \tau^+\tau^-$ pairs, while a 1.5 cm beampipe, a 4-layer
silicon detector and a small-cell inner drift chamber are used to record  
the remaining 189 $\times 10^6\ \tau^+\tau^-$ pairs~\cite{natk}.  
The detector is described in detail elsewhere~\cite{bel}. 

\section{Selection of $\tau^+\tau^-$ events}

We select events in which one $\tau$ decays 
to leptons, $\tau^-\to l^-\bar{\nu_{l}}\nu_{\tau},\ l=e,\mu$, 
while the other one decays via the hadronic channel 
$\tau^-\to h^-\nu_{\tau}$, where $h^-$ denotes the hadronic system.
%DRM Events where both $\tau$ decay to leptons are used for normalization.
Events where both $\tau$'s decay to leptons are used for normalization.
This reduces systematic uncertainties substantially.

The selection process, which is optimized to suppress background
while retaining a high efficiency for the decays under study,
proceeds in two stages. The criteria of the  first stage suppress beam background
to a negligible level and reject most of the background from other 
physical processes. These criteria retain a $46.0$\% efficiency 
for $\tau^+\tau^-$ events.
We then select events having $2$ to $4$ tracks with a net charge
less than or equal to one in absolute value. The extrapolation of each track 
to the interaction point (IP) is required to pass within $\pm 0.5\ \rm{cm}$ in the 
transverse direction and $\pm 2.5\ \rm{cm}$ in the longitudinal
direction of the nominal collision point of the beams. 
Each track must have a transverse momentum in the   center-of-mass (CM) frame 
larger than $0.1\ {\rm GeV}/c$.
At least one of the charged particles should have a transverse momentum higher
than $0.5\ {\rm GeV}/c$.
The sum of the absolute values of the CM track momenta must be less 
than $9\ {\rm GeV}/c$. The minimum opening angle for any pair of 
tracks is required to be larger than $20^{\rm o}$.
The number of photons with a CM energy exceeding $80\ {\rm MeV}$ is required to be less 
than or equal to five. The total ECL energy deposition in the laboratory frame
must be less than $9\ \rm{GeV}$. The total energy of all photon candidates in the 
laboratory frame should satisfy $\sum E_{\rm{\gamma}}^{\rm{LAB}}< 0.2\ {\rm GeV}$.
The missing four-momentum $P_{{\rm miss}}$ is calculated by 
subtracting the four-momentum of all charged tracks and photons 
from the beam four-momentum. The missing mass 
$M_{{\rm miss}}=\sqrt{P^2_{{\rm miss}}}$ is required to satisfy 
$1\ {\rm GeV}/c^2 \leq M_{{\rm miss}}\leq 7\ {\rm GeV}/c^2$. 
The polar angle of the missing momentum 
in the CM frame is required to be larger than or equal to $30^{\rm{o}}$ 
and less than or equal to $150^{\rm{o}}$. 
The last two criteria are particularly effective in suppressing the 
backgrounds from radiative Bhabha, $e^+ e^-\to \mu^+\mu^-(\gamma)$ and two-photon 
processes.

 At the second stage, two event classes are selected for further
processing: a two-lepton sample
$(l_1^{\pm},l_2^{\mp}),\ l_1,l_2=e,\mu$ and a lepton-hadron 
sample $(l^{\pm},K_S\pi^{\mp}),\ l=e,\mu$.
To select electrons, a likelihood ratio requirement 
$\mathcal{P}_{e}=\mathcal{L}_{e}/(\mathcal{L}_{e}+\mathcal{L}_{x})>0.8$ 
is applied, where the electron likelihood function $\mathcal{L}_{e}$ 
and the non-electron function $\mathcal{L}_{x}$ include information on 
the specific ionization ($dE/dx$) measurement by the CDC, the ratio of 
the cluster energy in the ECL to the track momentum 
measured in  the CDC, the transverse ECL shower shape and the light
yield in the ACC \cite{eid}.
The efficiency of this requirement for electrons is $93\%$.
To select muons, a likelihood ratio requirement
$\mathcal{P}_{\mu}=\mathcal{L}_{\mu}/(\mathcal{L}_{\mu}+\mathcal{L}_{\pi}+\mathcal{L}_{K})>0.8$ 
is applied. It provides $88\%$ efficiency for muons. 
Each of the muon($\mathcal{L}_{\mu}$), pion($\mathcal{L}_{\pi}$) and 
kaon($\mathcal{L}_{K}$) likelihood functions is evaluated from the information
on the difference between 
the range calculated from the momentum of the particle and the range
measured by KLM, and the $\chi^2$ of the KLM hits with respect 
to the extrapolated track \cite{muid}.
To separate pions from kaons, for each track we determine the pion 
($\mathcal{L'}_{\pi}$) and kaon ($\mathcal{L'}_{K}$) likelihoods from
the ACC response, the $dE/dx$ measurement in the CDC and the TOF
flight-time measurement, and form a likelihood ratio 
$\mathcal{P}_{K/\pi}=\mathcal{L'}_{K}/(\mathcal{L'}_{\pi}+\mathcal{L'}_K)$.
For pions we apply the requirement $\mathcal{P}_{K/\pi}<0.3$, which 
provides a pion identification efficiency of about $93\%$, while keeping 
the pion fake rate at the $6\%$ level.

To evaluate the background and to calculate efficiencies, a 
Monte Carlo (MC) sample of $1.50\times10^9~\tau^+\tau^-$ pairs 
is produced with the KORALB/TAUOLA generators \cite{koral,tola}. The 
detector response is simulated by a GEANT3 based program \cite{geant}. 

\subsection{Two-lepton events} 

For this class 
the $(e,e)$ and $(\mu,\mu)$ samples still contain 
contamination from radiative Bhabha and $e^+ e^-\to \mu^+\mu^-(\gamma)$ processes of about 50\%, 
only $(e,\mu)$ events are used for normalization.
To further suppress $B\bar{B}$ and charm backgrounds, we 
require the opening angle of the leptons to be larger than
$90^{\rm{o}}$ in the CM. 
As a result, we selected {\bf $2,018,000$} events of the $(e^+,\mu^-)$ type
and {\bf $2,028,000$} $(e^-,\mu^+)$ events.

MC simulation indicates that there is 
an approximately $5\%$ contamination 
coming primarily from the two-photon process 
$e^+ e^- \to e^+ e^- \mu^+ \mu^-$ ($2.0\%$) and 
from $\tau^+\tau^-\to e^+(\mu^+) \pi^-
\nu_{e}(\nu_{\mu})\nu_{\tau}\bar{\nu_{\tau}}$ 
events where the $\pi$ is misidentified as a lepton ($2.8\%$).  Contamination from other 
non-$\tau^+\tau^-$ processes is found to be negligible (less than 0.1\%).
The numbers of $(e^+,\mu^-)$ and $(e^-,\mu^+)$ events after background 
subtraction are $1,929,300\pm 1,400$ and $1,911,700\pm 1,400$, respectively. 
The detection efficiencies and their statistical errors are 
$(19.262\pm0.006)\%$ for $(e^+,\mu^-)$ and $(19.252\pm 0.006)\%$ 
for $(e^-,\mu^+)$ events.

\subsection{Lepton-hadron events}

\begin{figure}[htb]
\includegraphics[width=0.48\textwidth]{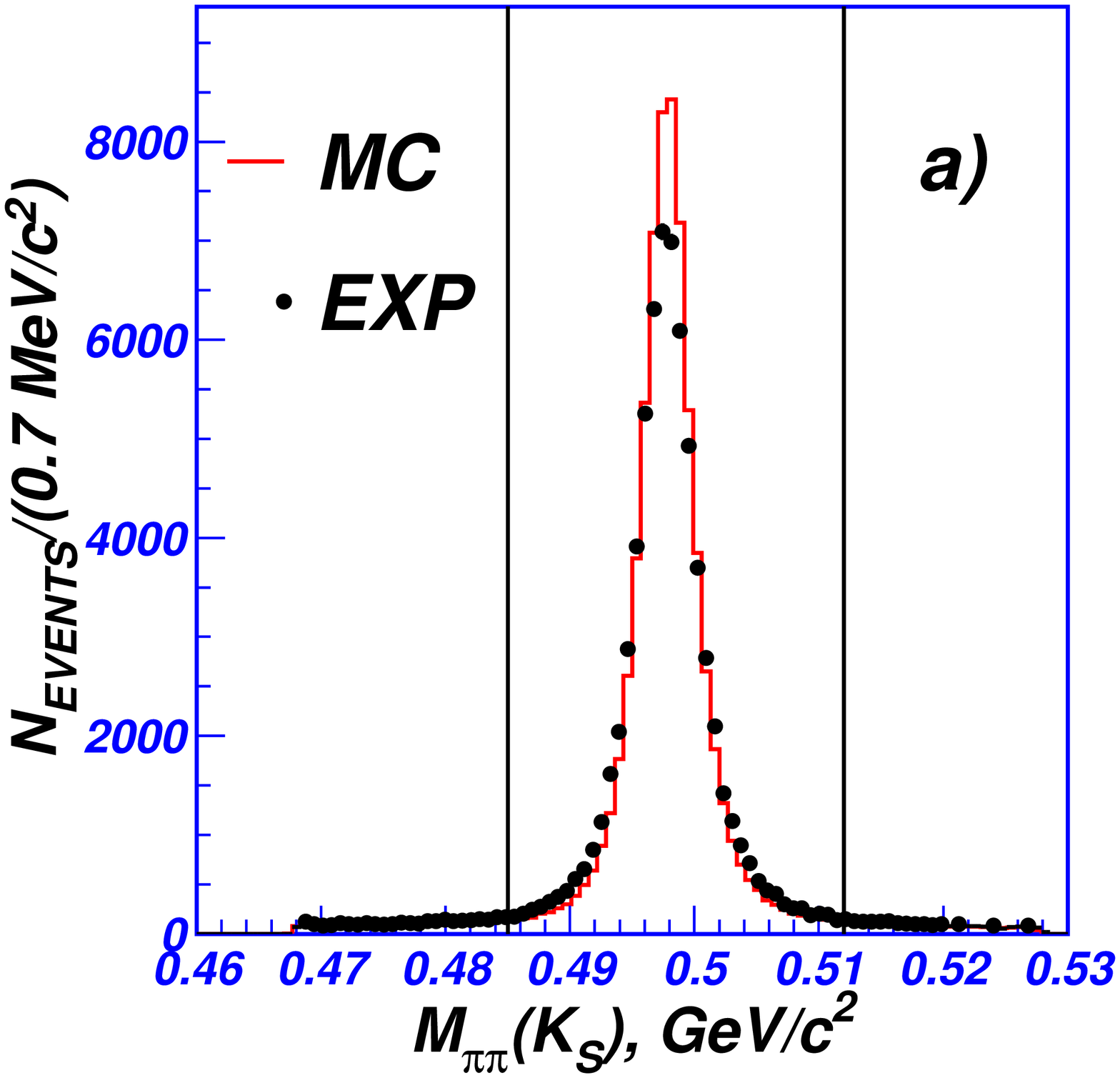}
\hfill
\includegraphics[width=0.48\textwidth]{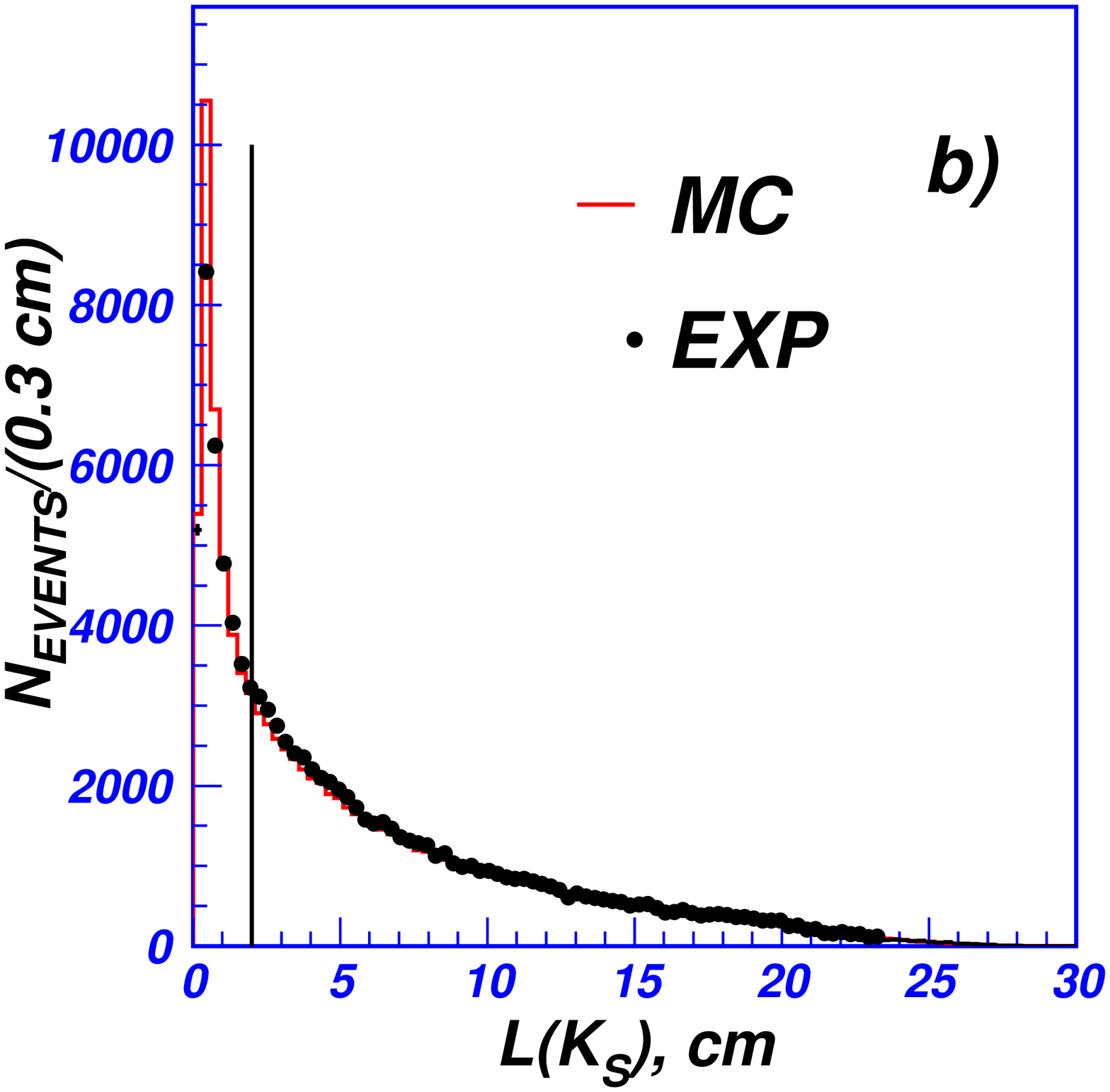}
\\
\parbox[t]{0.95\textwidth}{\caption{MC (histogram) and
    experimental data (points) distributions normalized to 
    the same number of events.
    (a) shows the $\pi^+\pi^-$ invariant mass distribution 
        for $K_S$ candidates.
    (b) shows the $K_S$ candidate decay length. 
    For each distribution all the criteria described in the 
    text except the one pertaining to the displayed parameter are applied. 
    Applied cuts are shown by vertical lines.\label{ksdiss}}}
\end{figure}

For this class we select events with only one lepton $l^{\pm}\ (l=e,\ \mu)$, one
$K_S$ candidate and one charged pion $\pi^{\mp}$.
A $K_S$ meson is reconstructed from a pair of oppositely charged pions 
having invariant mass $M_{\pi\pi}(K_S)$ within $\pm13.5$ MeV of the 
$K_S$ mass, which corresponds to a $\pm5\sigma$ signal range.
The pion momenta are then refitted with a common 
vertex constraint. The $z$-distance between the two helices at the 
$\pi^+\pi^-$ vertex position before the fit is required to be less
than $1.5\ {\rm cm}$, 
where $z$ is defined as the direction opposite to the positron beam.
The closest approach of at least one track to the IP 
in the $r-\varphi$ plane must be larger than $0.03\ {\rm cm}$.
The decay length of the $K_S$ candidate in the $r-\varphi$ plane must satisfy
$0.1\ {\rm cm} \leq L_{\rm{\perp}}\leq 20\ {\rm cm}$. The $z$-projection 
of the $K_S$ candidate decay length is required to be $L_{\rm{z}}\leq 20\ {\rm cm}$.
The $K_S$ decay length $L(K_S)=\sqrt{L^2_{\rm{\perp}}+L^2_{\rm{z}}}$ 
must be larger than $2\ {\rm cm}$. The cosine of the azimuthal angle between 
the momentum vector and the decay vertex vector of the  $K_S$ candidate 
is required to be larger than or equal to $0.95$. 
The lepton-$K_S$ and lepton-$\pi$ opening 
angles are required to be larger than $90^{\rm o}$ in the CM. 
$68,107$ events were selected for further analysis.
Figure \ref{ksdiss} shows a comparison of the MC 
and experimental distributions for the $\pi^+\pi^-$ invariant
mass of the $K_S$ candidate and the $K_S$ decay length.

\begin{figure}[htb]
\includegraphics[width=0.48\textwidth]{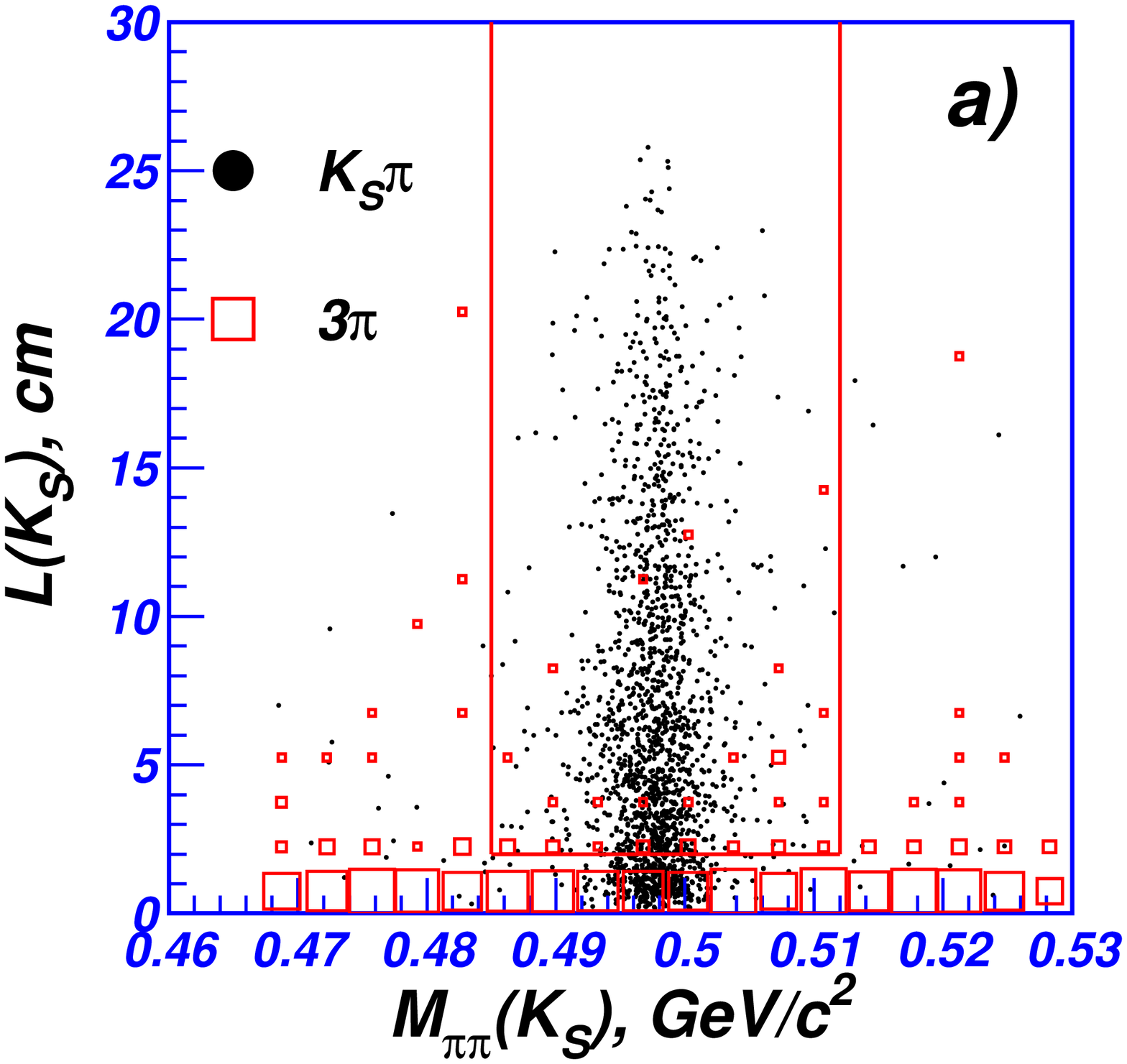}
\hfill
\includegraphics[width=0.48\textwidth]{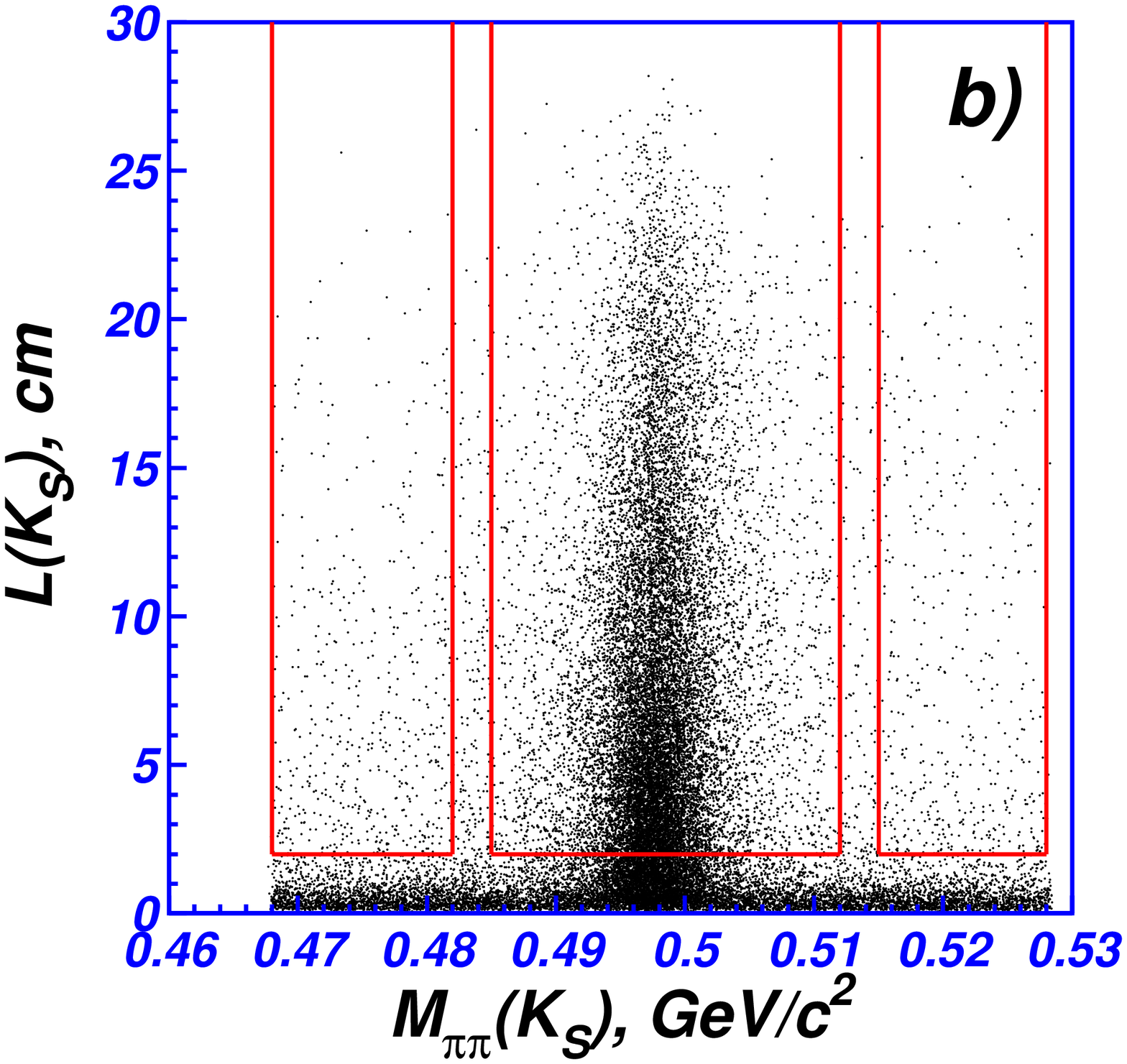}
\\
\parbox[t]{0.95\textwidth}{\caption{Decay length {\it vs.}
 $\pi\pi$ invariant mass of the $K_S$ candidate 
 for $(e^{+}, K_S \pi^{-})$ events.
 All selection criteria described in the text except for 
 those pertaining to the parameters being displayed are applied.  
(a) shows MC data, where events with a real $K_S$ are plotted as points, and 
    the events with fake $K_S$'s from
    $\tau^-\to\pi^+\pi^-\pi^-\nu_{\tau}$ are plotted as boxes, whose sizes are proportional to the
    number of entries.    
(b) shows experimental data. The signal region is indicated by 
    the middle rectangle, while sideband regions are shown by 
    the rectangles to the left and right of the signal region.\label{lmks}}}
\end{figure}

{\noindent Figure~\ref{ksdiss} (a) shows that MC $\pi^+\pi^-$ mass resolution is
slightly better than the experimental one resulting in a clear
difference of the $\pi^+\pi^-$ mass spectra in the region of the $K_S$ peak.
However, the efficiency of the $M_{\pi\pi}$ cut for the $K_S$ candidates
is almost $100\%$, hence the impact of this discrepancy on the
detection efficiency is very small and is taken into account in the
systematic uncertainty.
In Fig.~\ref{ksdiss} (b) one can see a clear difference between the $L(K_S)$
distributions in the region of small $L(K_S)$, where
events of $\tau^-\to\pi^-\pi^-\pi^+\nu_{\tau}$ decay
are located, however, in the region, where $L(K_S)>2$~cm,
populated mostly by true $K_S$'s the agreement is good.
Figure \ref{lmks} shows selected events on a plot of the $K_S$ decay 
length versus the $\pi^+\pi^-$ invariant mass of the $K_S$ candidate.}
The main background is from other $\tau$ decays: 
$\tau^-\to K_S \pi^- K_L \nu_{\tau}$, 
$\tau^-\to K_S \pi^- \pi^0 \nu_{\tau}$, 
$\tau^-\to K_S K^- \nu_{\tau}$, 
$\tau^-\to \pi^- \pi^- \pi^+ \nu_{\tau}$. 
Using the branching fractions of these decays from Ref.~\cite{pdg} 
and detection efficiencies from MC simulation, the contamination 
from decays with a $K_S$ is calculated to be $14.7\%$. 
$\tau^- \to \pi^- \pi^- \pi^+ \nu_{\tau}$ decays 
contaminate the sample 
when a pair of oppositely charged pions is reconstructed as a fake $K_S$.
The $\pi^+\pi^-$ invariant mass distribution 
of these fake $K_S$'s is flat in the region of the $K_S$ mass (see also 
Fig.~\ref{lmks}). The number of $3\pi$ background events is calculated from two 
sideband regions in the $L(K_S)$ {\it vs.} $M_{\pi\pi}(K_S)$ plane, 
determined by the following criteria: $468\ {\rm MeV}/c^2<
M_{\pi\pi}(K_S)<482\ {\rm MeV}/c^2$ and $L(K_S)>2\ {\rm cm}$ 
for the first region, $515\ {\rm MeV}/c^2 <M_{\pi\pi}(K_S)<528\ 
{\rm MeV}/c^2$ and $L(K_S)>2\ {\rm cm}$ 
for the second one. These sidebands have the same area as the signal region.  
The fraction of signal events in the $3\pi$-sideband region is 
about $1\%$, which is taken into account in the calculation of the 
MC signal detection efficiency. We observe a $5.6\%$ background of $3\pi$
events in the signal region.
In the $(l^{\pm},K_S\pi^{\mp}),\ l=e,\mu$ sample there is a small contamination 
(of about $0.3\%$ for the $e$-tagged and $2.4\%$ for the $\mu$-tagged
events) coming primarily from $(\pi^{\pm},K_S\pi^{\mp})$
events, where the first pion was misidentified as a lepton. 
The non-$\tau^+\tau^-$ background is found to be $0.6\%$.
After background subtraction $53,110\pm 271$ signal events remain. Table~\ref{all}
shows how they are distributed among the various tagging configurations.

\begin{table}[htb]
\caption{Branching fractions for different tagging configurations}
\label{all}
\begin{tabular}
{l|c|c|c|c}
\hline
  & $(e^+,K_S\pi^-)  $ & $(e^-,K_S\pi^+)  $ %
  & $(\mu^+,K_S\pi^-)$ & $(\mu^-,K_S\pi^+)$ \\
\hline
$N_{\rm{exp}}$     & $13336\pm 137$  & $13308\pm 137$  % 
                   & $13230\pm 134$  & $13236\pm 134$  \\ 
$\varepsilon(l,K_S\pi),\%$ & $5.70\pm 0.02$ & $5.58\pm 0.02$ & $5.95\pm 0.02$ & $5.89\pm 0.02$ \\

$\mathcal{B}(K_S\pi\nu),\%$ & $0.406\pm 0.005$ & $0.414\pm 0.005$ & $0.397\pm 0.005$ & $0.400\pm 0.005 $ \\
\hline
$<\mathcal{B}>_{l},\%$ & \multicolumn{2}{c|}{$0.410\pm 0.003$} &
  \multicolumn{2}{c}{$0.399\pm 0.003$} \\
\hline
$<\mathcal{B}>_{\rm all},\%$ & \multicolumn{4}{c}{$0.404\pm 0.002$} \\
\hline
\end{tabular} 
\end{table}

\section{$\tau^- \to K_S \pi^- \nu_{\tau}$ branching fraction}

The $\tau^- \to K_S \pi^- \nu_{\tau}$ branching fraction is calculated
according to the formula:

\begin{equation}
\mathcal{B}(K_S\pi^{\mp}\nu_{\tau})=
\frac{N(l_1^{\pm},K_S\pi^{\mp})}{N(l_1^{\pm},l_2^{\mp})}\cdot
\frac{\varepsilon(l_1^{\pm},l_2^{\mp})}
{\varepsilon(l_1^{\pm},K_S\pi^{\mp})}\cdot
\mathcal{B}(l_2^{\mp}\nu_{l}\nu_{\tau}),\ l_{1,2}=e,\mu, 
\label{ksone}
\end{equation}
where $N(l_1^{\pm},K_S\pi^{\mp})$,
 $\varepsilon(l_1^{\pm},K_S\pi^{\mp})$ are the number and MC efficiency
of the signal $(l_1^{\pm},K_S\pi^{\mp})$ events, 
$N(l_1^{\pm},l_2^{\mp})$, $\varepsilon(l_1^{\pm},l_2^{\mp})$ are
the number and MC efficiency of the two-lepton $(l_1^{\pm},l_2^{\mp})$ 
events, $\mathcal{B}(l_2^{\mp}\nu_{l}\nu_{\tau})$ is the $\tau$ leptonic 
branching fraction taken from Ref.~\cite{pdg}.
Note that the tag-lepton ($l_1^{\pm}$) efficiency cancels in the ratio of the 
efficiencies, so the associated systematic uncertainty is reduced. 
The branching fractions calculated separately  for each event
configuration are given in Table~\ref{all}, which also lists 
separately the averages for electrons and muons as well as the
overall branching fraction. 

\begin{table}[htb]
\caption{Systematic uncertainties}
\label{tsys}
\begin{tabular}
{l|c}
\hline
Source & Contribution,\% \\
\hline
$K_S$ detection efficiency            & $2.5$ \\
$\tau^+\tau^-$ background subtraction     & $1.6$ \\
$\sum E^{\rm{LAB}}_{\rm{\gamma}}$     & $1.0$ \\
Lepton identification efficiency      & $0.8$ \\
Pion momentum                         & $0.5$ \\
Non-$\tau^+\tau^-$ background subtraction & $0.3$ \\
$\mathcal{B}(l\nu_{l}\nu_{\tau})$     & $0.3$ \\
$\frac{\varepsilon(l_1,l_2)}{\varepsilon(l_1,K_S\pi)}$ & $0.2$ \\
$K_S$ momentum                        & $0.2$ \\
Pion identification efficiency        & $0.1$ \\
\hline
Total                                 & $3.3$ \\ 
\hline
\end{tabular}
\end{table}

Table \ref{tsys} lists the different sources of systematic uncertainties 
for the branching fraction. The dominant contributions come from the 
$K_S$ detection efficiency and background subtraction. 
A systematic uncertainty in the $K_S$ detection efficiency receives
contributions from the reconstruction of $K_S$ daughter
pions ($2.3\%$), the efficiency for fitting two
pion tracks to a common $\pi^+\pi^-$ vertex ($0.9\%$), 
which was evaluated by varying the cut on the $z$-distance between the two 
helices at the vertex position before the fit, 
and the efficiency of the selection criteria ($0.6\%$), which was checked by
varying cuts on the $\pi^+\pi^-$ invariant mass $M_{\pi\pi}(K_S)$. 
Systematic uncertainties arising from $\tau^+\tau^-$-background subtraction 
are $0.8\%$, $1.1\%$, $0.6\%$ and $0.5\%$ for the 
$\tau^- \to K_S K_L \pi^- \nu_{\tau}$, $\tau^- \to K_S \pi^- \pi^0 \nu_{\tau}$, 
$\tau^- \to K_S K^- \nu_{\tau}$ and $\tau^- \to \pi^- \pi^- \pi^+ \nu_{\tau}$ 
modes, respectively. For the background from $\tau$ decay modes
with a $K_S$ the uncertainties are determined by the corresponding uncertainties in
their branching fractions taken from Ref.~\cite{pdg}, except for the 
$\tau^- \to K_S K_L \pi^- \nu_{\tau}$ mode. Here we
rely on the isospin relation $\mathcal{B}(\tau^-\to K_S
K_L\pi^-\nu_{\tau})=1/2\mathcal{B}(\tau^-\to K^+K^-\pi^-\nu_{\tau})$ 
and the CLEO result~\cite{clf} to calculate the $\tau^-\to K_S K_L \pi^-\nu$
branching fraction 
$\mathcal{B}(\tau^- \to K_S K_L \pi^- \nu_{\tau})=(0.078\pm0.006)\%$. 
The uncertainty in the contamination by $\tau^- \to \pi^- \pi^- \pi^+ \nu_{\tau}$ 
events is evaluated by varying the $K_S$ decay length cut. 

The lepton detection efficiency is corrected using the 
$e^+e^-\to e^+e^- l^+ l^-,\ l=e,\mu$ two-photon data sample. 
An efficiency correction table is calculated in 70 bins
on the plane of momentum {\it vs.} polar angle in the laboratory frame and then 
applied to the Monte Carlo efficiencies 
$\varepsilon(l_1^{\pm},K_S\pi^{\mp})$ and $\varepsilon(l_1^{\pm},l_2^{\mp})$. 
Hence, the uncertainty on the leptonic efficiency is determined by the 
statistics of  the $e^+e^-\to e^+e^- l^+ l^-$ sample and
the long-term stability, which is evaluated from the variation 
of the corrections calculated for time ordered subsamples of the 
experimental two-photon data.
The pion identification efficiency in MC differs from that in data.
In the signal sample, $K_S$ mesons provide a 
source of identified pions, which are used to calculate corrections to the
MC efficiency. Therefore, the systematic 
uncertainty on the pion identification efficiency is determined 
by the statistical error of the correction, which is about $0.1\%$.
To calculate $\varepsilon(l_1,K_S\pi)$ a signal MC sample is produced 
according to the $K^*(892)+K^*(1680)$ model and the model dependence of 
$\varepsilon(l_1,K_S\pi)$ is found to be negligible. 

We also vary cuts on the pion momentum, the kaon momentum, and the total 
laboratory energy of photons ($\sum E^{\rm{LAB}}_{\rm{\gamma}}$) to check
the stability of the branching fraction.
The total systematic uncertainty of $3.3\%$ is obtained by adding all the
contributions in quadrature. Our final result for the branching
fraction is $\mathcal{B}(\tau^-\to K_S\pi^-\nu_{\tau})=
(0.404\pm 0.002({\rm stat.})\pm 0.013({\rm syst.}))\%$.

\section{Analysis of the $\tau^-\to K_S\pi^-\nu_{\tau}$ spectrum}

\begin{figure}[htb]
\includegraphics[width=0.508\textwidth]{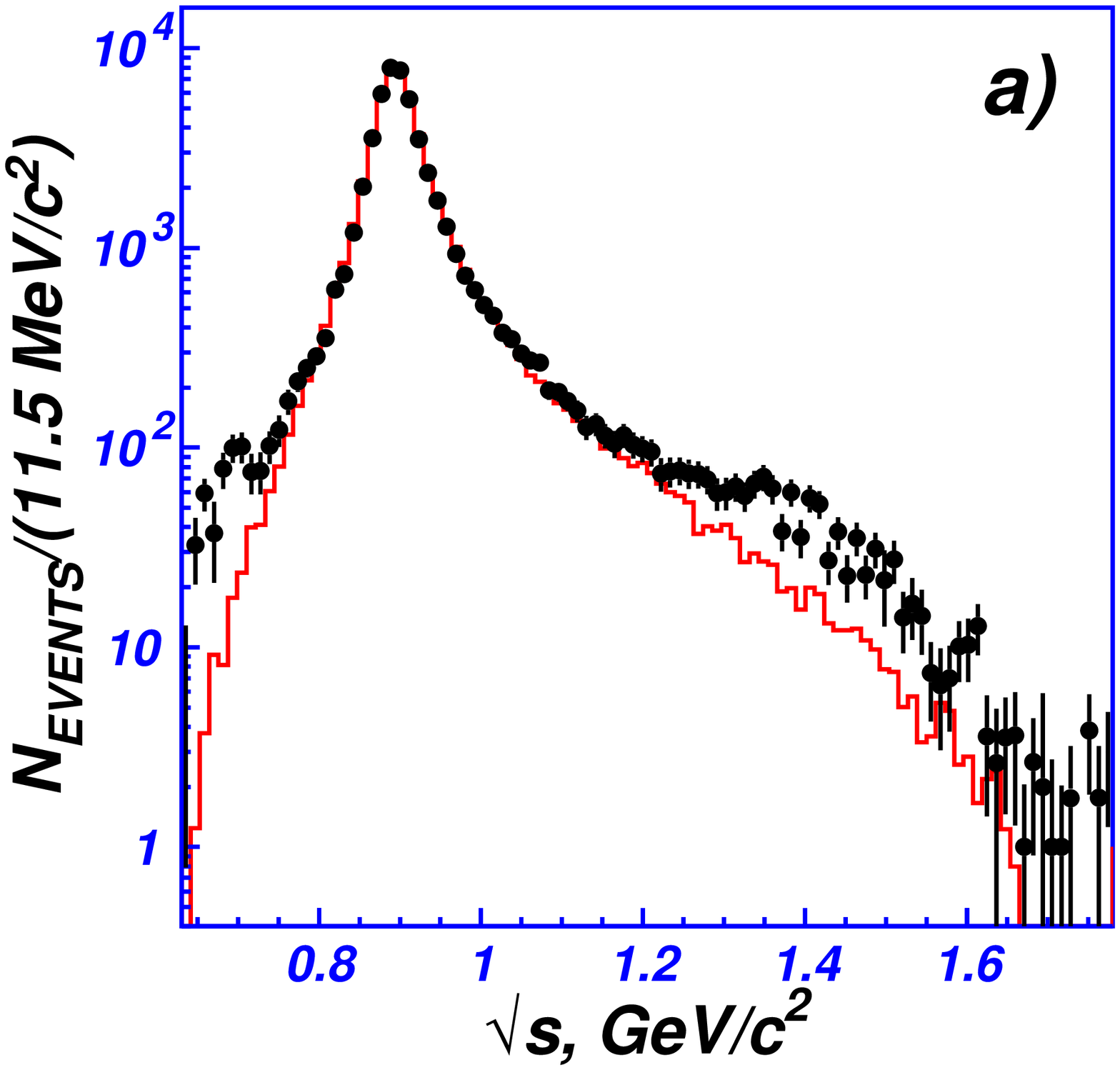}
\hfill
\includegraphics[width=0.48\textwidth]{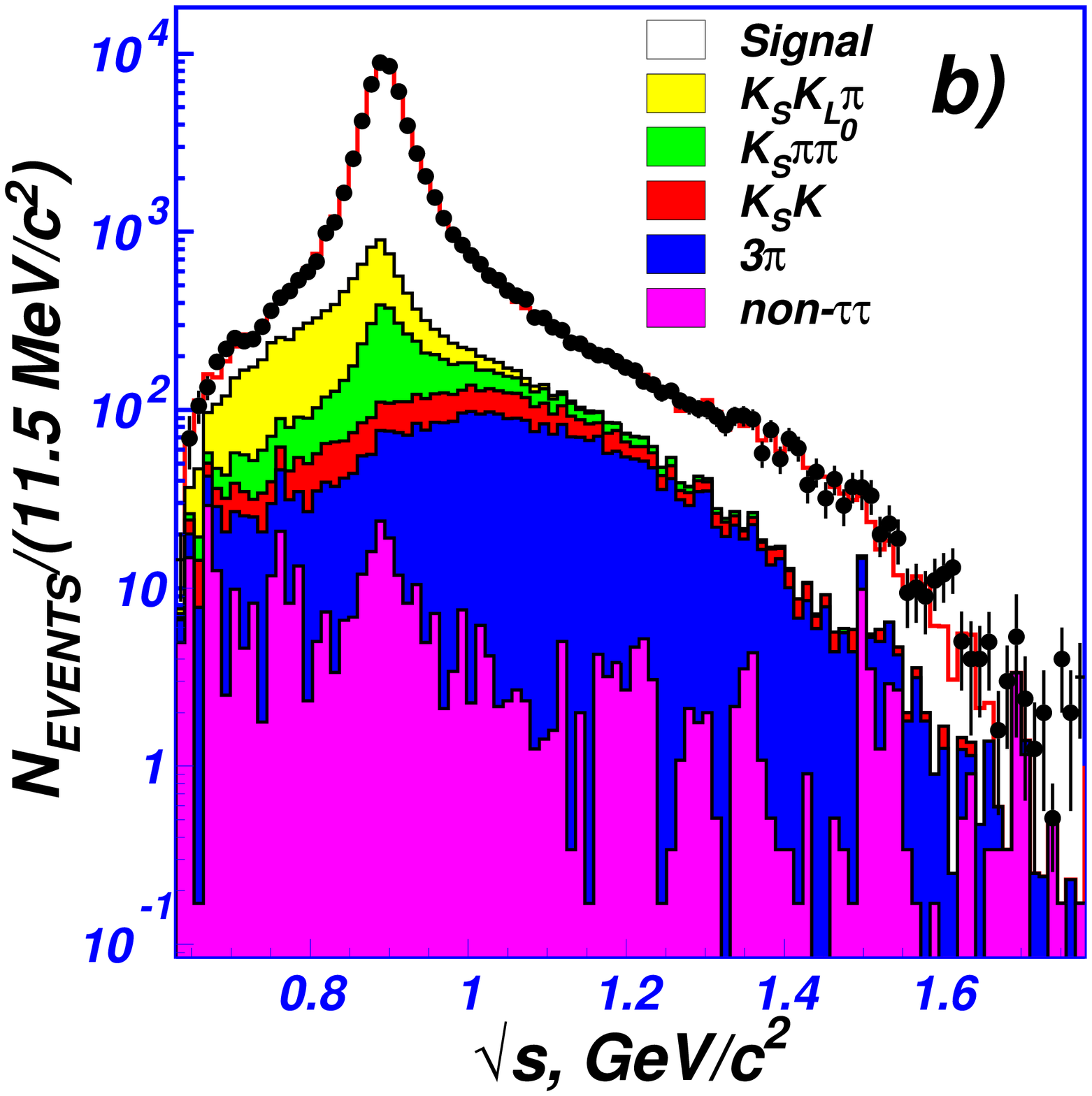}
\\
\parbox[t]{0.95\textwidth}{\caption{ Comparison of the $K_S\pi$ 
mass distributions, points are experimental data, histograms are 
spectra expected for different models.
(a) shows the fitted result with the model 
    incorporating the $K^*(892)$ alone, here the background 
    has been already subtracted from both experimental
    and expected distributions. 
(b) shows the fitted result with the
    $K^*(892)+K_0^*(800)+K^*(1410)$ model, here
    different types of background are also shown.\label{ksto}}}
\end{figure}

The $K_S \pi^-$ invariant mass distribution shown in  
Fig.~\ref{ksto} exhibits a very clear $K^*(892)^-$ signal. 
We parameterize this spectrum  by the following function 
(see Ref.~\cite{fione} for more detail):
\begin{equation}
\frac{d\Gamma}{d\sqrt{s}}\sim
\frac{1}{s}\biggl(1-\frac{s}{m^2_{\tau}}\biggr)^2 \biggl(1+2\frac{s}{m^2_{\tau}}\biggr)P\biggl\{P^2|F_V|^2+\frac{3(m^2_K-m^2_{\pi})^2}{4s(1+2\frac{s}{m^2_{\tau}})}|F_S|^2 \biggr\}, 
\label{dra}
\end{equation}
{\noindent where $s$ is the $K_S\pi^-$ invariant mass squared and $P$ is the 
$K_S$ momentum in the $K_S\pi^-$ rest frame:}

\begin{equation}
P(s)=\frac{1}{2\sqrt{s}}\sqrt{\bigl[s-(m_K+m_{\pi})^2\bigr]\bigl[s-(m_K-m_{\pi})^2\bigr]}.\label{mom}
\end{equation}

The vector form factor $F_V$ is parameterized by the $K^*(892)$,
$K^*(1410)$ and $K^*(1680)$ meson amplitudes:
\begin{equation}
F_V=\frac{1}{1+\beta+\chi}\biggl[BW_{K^*(892)}(s)+\beta
BW_{K^*(1410)}(s)+\chi BW_{K^*(1680)}(s)\biggr],
\end{equation}
{\noindent where $\beta$ and $\chi$ are complex coefficients
for the fractions of the $K^*(1410)$ and $K^*(1680)$ resonances, respectively.
$BW_{R}(s),\ (R=K^*(892),\ K^*(1410),\ K^*(1680))$ is a
relativistic Breit-Wigner function:}

\begin{equation}
BW_{R}(s)=\frac{M_R^2}{s-M_R^2+i\sqrt{s}\Gamma_{R}(s)}, 
\end{equation}
{\noindent where $\Gamma_{R}(s)$ is the s-dependent total width of the resonance:}

\begin{equation}
\Gamma_{R}(s)=\Gamma_{0R}\frac{M_R^2}{s}\biggl( \frac{P(s)}{P(M_R^2)}\biggr)^{2\ell+1},
\end{equation}
{\noindent where $\ell=1(0)$ if the $K\pi$ system originates in the $P(S)$-wave
state and $\Gamma_{0R}$ is the resonance width at its peak.}
  
The scalar form factor $F_S$ includes the $K_0^*(800)$ and
$K_0^*(1430)$ contributions, their fractions are described respectively by 
the complex constants $\varkappa$ and $\gamma$:  

\begin{equation}
F_S=\varkappa\frac{s}{M_{K_0^*(800)}^2}BW_{K_0^*(800)}(s)+
\gamma\frac{s}{M_{K_0^*(1430)}^2}BW_{K_0^*(1430)}(s).
\end{equation}

The experimental distribution is approximated in the mass range 
from 0.63~GeV/$c^2$ to 1.78~GeV/$c^2$
by a function calculated from the convolution of the spectrum given
by Eq.~(\ref{dra}) and the detector response function, which takes into account 
the efficiency and finite resolution of the detector. 
In all fits the $K^*(892)$ mass and width as well as the total 
normalization are free parameters. Only the strengths (fractions) of 
the other $K^*$'s are free parameters, while their masses and widths
are fixed at the world average values~\cite{pdg}. 
In the approximation $\varkappa$ is chosen to be real, because $F_S$ 
is defined up to the common phase, which cancels in $|F_S|^2$.

\begin{table}[htb]
\caption{Results of the fit of the $K_S\pi$ mass spectrum in
different models of the  non-$K^*(892)$ mechanism: the $K^*(1410)$ 
and  $K^*(1680)$ contributions are described by the 
complex constants $\beta$ and $\chi$, respectively, while
that from the $K_0^*(800)$ is described by the real constant $\varkappa$. 
Masses and widths of the non-$K^*(892)$ resonances were fixed at their PDG
values (the $K_0^*(800)$ mass and width were fixed from Ref. \cite{abli}).}
\label{tfit}
\begin{tabular}
{l|c|c|c}
\hline
 & $K^*(892)$ & $K_0^*(800)+K^*(892)+$ & $K_0^*(800)+K^*(892)+$ \\
 &            & $+K^*(1410)$           & $+K^*(1680)$           \\
\hline
$M_{K^*(892)^-},\ {\rm MeV}/c^2$  & $895.53 \pm 0.19$ & $895.47\pm 0.20$ & $894.88 \pm 0.20$ \\
$\Gamma_{K^*(892)^-},\ {\rm MeV}$ & $49.29 \pm 0.46$  & $46.19\pm 0.57$  & $45.52 \pm 0.51$  \\
$|\beta|$                      &     & $0.075\pm 0.006$ &                \\
$\arg(\beta)$                   &     & $ 1.44\pm 0.15 $ &                \\
$|\chi|$                       &     &                  & $0.117\pm \begin{array}{l} 0.017 \\ 0.033 \end{array}$ \\
$\arg(\chi)$                    &     &                  & $3.17\pm 0.47$ \\ 
$\varkappa$                    &     & $1.57\pm 0.23$   & $1.53\pm 0.24$ \\
\hline
$\chi^2/{\rm n.d.f.}$                   & $448.4/87$ & $90.2/84$        & $106.8/84$     \\
$P(\chi^2),\%$                 & $0$ & $30$             & $5$            \\
\hline
\end{tabular} 
\end{table}

Figure~\ref{ksto} (a) and Table \ref{tfit} show that 
the $K^*(892)$ alone is 
not sufficient to describe the $K_S\pi$ mass spectrum. 
To describe the enhancement near threshold, 
we introduce a $K_0^*(800)$ amplitude, while  
for description of the distribution at higher invariant 
masses we try to include the $K^*(1410)$, $K^*(1680)$ 
vector resonances (see Table \ref{tfit}) or the scalar $K_0^*(1430)$ 
(see Table \ref{tfitt}). Figure~\ref{ksto} (b) demonstrates the good
quality of the fit with the $K_0^*(800)+K^*(892)+K^*(1410)$ model. 
It can be seen from Tables \ref{tfit}, \ref{tfitt}
that we cannot distinguish between the $K_0^*(800)+K^*(892)+K^*(1410)$ and 
$K_0^*(800)+K^*(892)+K_0^*(1430)$ models. 
The fit quality with the $K_0^*(800)+K^*(892)+K^*(1680)$ model  (see
the fourth column of Table \ref{tfit}) is worse than that of the 
$K_0^*(800)+K^*(892)+K^*(1410)$ and $K_0^*(800)+K^*(892)+K_0^*(1430)$ models.  

\begin{table}[htb]
\caption{Results of the fit of the $K_S\pi$ mass spectrum in
the $K_0^*(800)+K^*(892)+K^*_0(1430)$ model (two solutions). The $K_0^*(1430)$
contribution is described by the complex constant $\gamma$, while that from the
$K_0^*(800)$ is described by the real constant $\varkappa$. 
Masses and widths of the non-$K^*(892)$ resonances were fixed at their PDG
values (the $K_0^*(800)$ mass and width were fixed from Ref. \cite{abli}).}
\label{tfitt}
\begin{tabular}
{l|c|c}
\hline
   & \multicolumn{2}{c}{$K_0^*(800)+K^*(892)+K^*_0(1430)$} \\
\cline{2-3}
&  solution 1 & solution 2 \\
\hline
$M_{K^*(892)^-},\ {\rm MeV}/c^2$  & $895.42\pm 0.19$ & $895.50\pm 0.22$ \\
$\Gamma_{K^*(892)^-},\ {\rm MeV}$ & $46.14\pm 0.55$  & $ 46.20\pm 0.69$ \\
$|\gamma|$                        & $0.954\pm 0.081$ & $  1.92\pm 0.20$ \\
$\arg(\gamma)$                    & $0.62\pm 0.34$   & $  4.03\pm 0.09$ \\
$\varkappa$                       & $1.27\pm 0.22$   & $  2.28\pm 0.47$ \\
\hline
$\chi^2/{\rm n.d.f.}$             & $86.5/84$        & $95.1/84$        \\  
$P(\chi^2),\%$                    & $41$             & $19$             \\
\hline
\end{tabular} 
\end{table}

\begin{table}[hbtp]
\caption{Results of the fit of the $K_S\pi$ mass spectrum in
the model when the non-$K^*(892)$ mechanism is introduced by the LASS 
scalar form factor, described by the parameters $a$ and $b$.}\label{tfittt}
\begin{tabular}{l|c|c}
\hline
   & $K^*(892)+$LASS & $K^*(892)+$LASS \\
   & $a$, $b$ - fixed     &  $a$, $b$ - free  \\ 
\hline
$M_{K^*(892)^-},\ {\rm MeV}/c^2$  & $895.42 \pm 0.19$ & $895.38\pm 0.23$ \\
$\Gamma_{K^*(892)^-},\ {\rm MeV}$ & $46.46 \pm 0.47$  & $46.53\pm 0.50$  \\
$\lambda$                       & $0.282\pm 0.011$  & $0.298\pm 0.012$ \\
$a,\  ({\rm GeV}/c)^{-1}$        & $2.13\pm 0.10$    & $10.9\pm \begin{array}{l} 7.4 \\ 3.0 \end{array}$ \\
$b,\ ({\rm GeV}/c)^{-1}$        & $3.96\pm 0.31$    & $19.0\pm \begin{array}{l} 4.5 \\ 3.6 \end{array}$ \\
\hline
$\chi^2/{\rm n.d.f.}$              & $196.9/86$        & $97.3/83$ \\  
$P(\chi^2),\%$                  & $10^{-8}$         & $13$      \\
\hline
\end{tabular} 
\normalsize
\end{table}

It should be noted that the absolute value of a sum of two Breit-Wigner 
functions of mass ($\sqrt{s}$) can have the same shape for two
different sets of parameters. In the case of the 
$K_0^*(800)+K^*(892)+K^*_0(1430)$ model the relevant parameters are $\varkappa$, 
$|\gamma|$ and $\arg(\gamma)$.
This statement holds true when mass-independent widths are considered. 
If the width is mass-dependent, some difference in the spectra appears.
If in the fit to the data the errors are large enough, we cannot 
distinguish these solutions by their $\chi^2$ values. For high
statistics the two solutions can be distinguished by a $\chi^2$ test. While for the 
$K_0^*(800)+K^*(892)+K^*(1410)$ and  $K_0^*(800)+K^*(892)+K^*(1680)$
models with a complicated vector form factor the values of $\chi^2$ 
are significantly different (due to the small ($\sim 1\%$) errors 
at the $K^*(892)$ peak), in the $K_0^*(800)+K^*(892)+K^*_0(1430)$ case
with a complicated scalar form factor different solutions result in
similar $P(\chi^2)$ values (see Table \ref{tfitt}) due to the relatively low statistics
in the region of the $K_0^*(800)$ and $K^*_0(1430)$ peaks.

An alternative way to describe our data is to use the 
parameterization of the scalar form factor suggested by the LASS 
experiment~\cite{lass,baba}: 
\begin{equation}
F_S=\lambda A_{{\rm LASS}}(s),\ A_{{\rm LASS}}=\frac{\sqrt{s}}{P}(\sin\delta_Be^{i\delta_B}+e^{2i\delta_B}BW_{K_0^*(1430)}(s)),
\end{equation}
where $\lambda$ is a real constant, $P$ is $K_S$ momentum in the
$K_S\pi$ rest frame (see Eq.~(\ref{mom})), and the phase  $\delta_B$ is 
determined from the equation 
$\cot\delta_B=\frac{1}{aP}+\frac{bP}{2}$, where $a$, $b$ are the model 
parameters. In this parameterization the non-resonant mechanism is 
given by the effective range term $\sin\delta_Be^{i\delta_B}$,
while the resonant structure is described by the $K_0^*(1430)$ amplitude.
  
Table \ref{tfittt} shows the results of fits to the spectrum in 
models, where the non-$K^*(892)$ mechanism is described by the
LASS parameterization of the scalar form factor. In the first fit
(see the second column of Table \ref{tfittt}) $a$ and $b$ parameters
were fixed at the LASS optimal values \cite{baba}. In the second fit
$a$ and $b$ were free parameters (see the third column of Table \ref{tfittt}). 
The optimal values of $a$ and $b$  in our fit  differ
significantly from the values obtained by the LASS collaboration in 
experiments on $K\pi$ scattering \cite{lass}.

The $K_0^*(800)+K^*(892)+K^*(1410)$ model was 
considered as the default and was used to obtain the $K^*(892)(K_S\pi)\nu$ fraction 
in the $K_S\pi\nu$ final state, which was found to be 
$\mathcal{B}(\tau^-\to K^*(892)^-\nu_{\tau})\cdot\mathcal{B}(K^*(892)^-\to K_S\pi^-)/\mathcal{B}(\tau^-\to K_S\pi^-\nu_{\tau})=0.933\pm 0.027$. 
The $0.027$ error includes the model uncertainty, 
which was found by calculating this fraction in the 
fits with the other models mentioned above, as well as the 
uncertainty in the fit parameters. Finally we obtain 
$\mathcal{B}(\tau^-\to K^*(892)^-\nu_{\tau})\cdot\mathcal{B}(K^*(892)^-\to K_S\pi^-)=(3.77\pm 0.02({\rm stat.})\pm 0.12({\rm syst.})\pm 0.12({\rm mod.}))\times 10^{-3}$.

\section{Measurement of the $K^*(892)^-$ mass and width}

A fit to the $K_S\pi^-$ invariant mass spectrum also provides 
a high precision measurement of the  $K^*(892)^-$ mass and width.
We consider a fit with the $K_0^*(800)+K^*(892)+K^*(1410)$ model,
which provides a good description of the data,
as a reference, and use it to obtain the $K^*(892)^-$ mass and width values. 
It can be seen from Table~\ref{tfit} that the statistical uncertainty
is about $0.20\ {\rm MeV}/c^2$ for the mass and $0.6\ {\rm MeV}$ for the width.
Two additional sources of uncertainty are studied: the effects of 
imperfect knowledge of the detector response function and model uncertainty.

The systematic uncertainty is studied with a MC sample by comparing 
the $K^*(892)^-$ parameters implemented in the generator and its parameters
after the full reconstruction procedure (the detector response
function is determined from other statistically independent MC
simulations of signal events). 
It is found to be $0.44\ {\rm MeV}/c^2$ for the mass and $1.0\ {\rm MeV}$
for the width.

The model uncertainty is investigated by fitting the $K_S\pi^-$ mass spectrum
with different models. The maximal difference from the reference value
is considered as a model uncertainty. It is found to be 
$0.59\ {\rm MeV}/c^2$ for the mass and $0.7\ {\rm MeV}$ for the width.

As a result, the $K^*(892)^-$ mass and width are 
$M(K^*(892)^-)=(895.47\pm 0.20({\rm stat.})\pm 0.44({\rm syst.})\pm
0.59({\rm mod.}))\ {\rm MeV}/c^2$  and $\Gamma(K^*(892)^-)=(46.2\pm
0.6({\rm stat.})\pm 1.0({\rm syst.})\pm 0.7({\rm mod.}))\ {\rm MeV}$,
where the first uncertainty is statistical, the second is systematic and the
third is from the model. 

\section{Conclusions}

\begin{figure}[htb]
\includegraphics[width=0.48\textwidth]{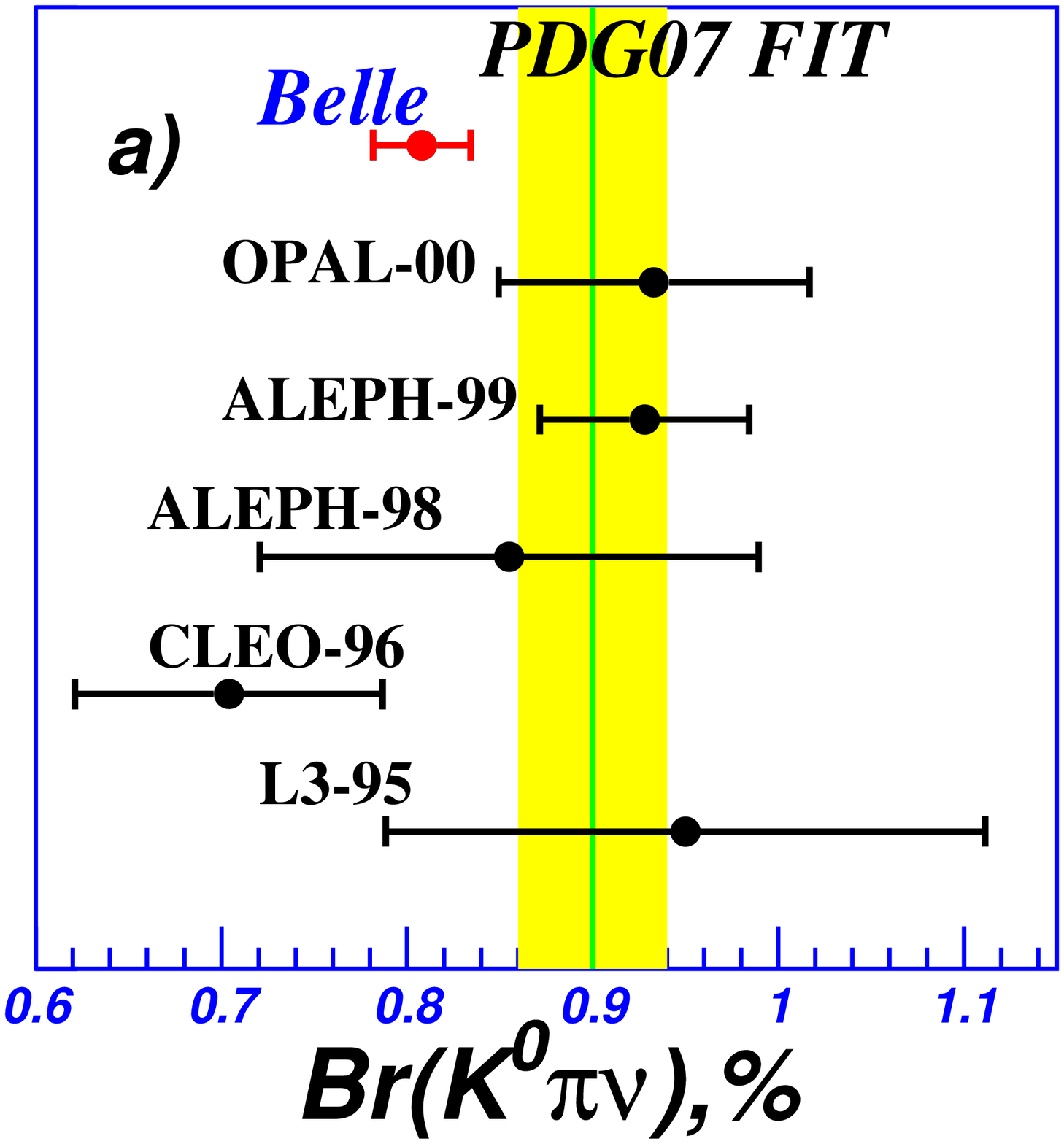}
\hfill
\includegraphics[width=0.48\textwidth]{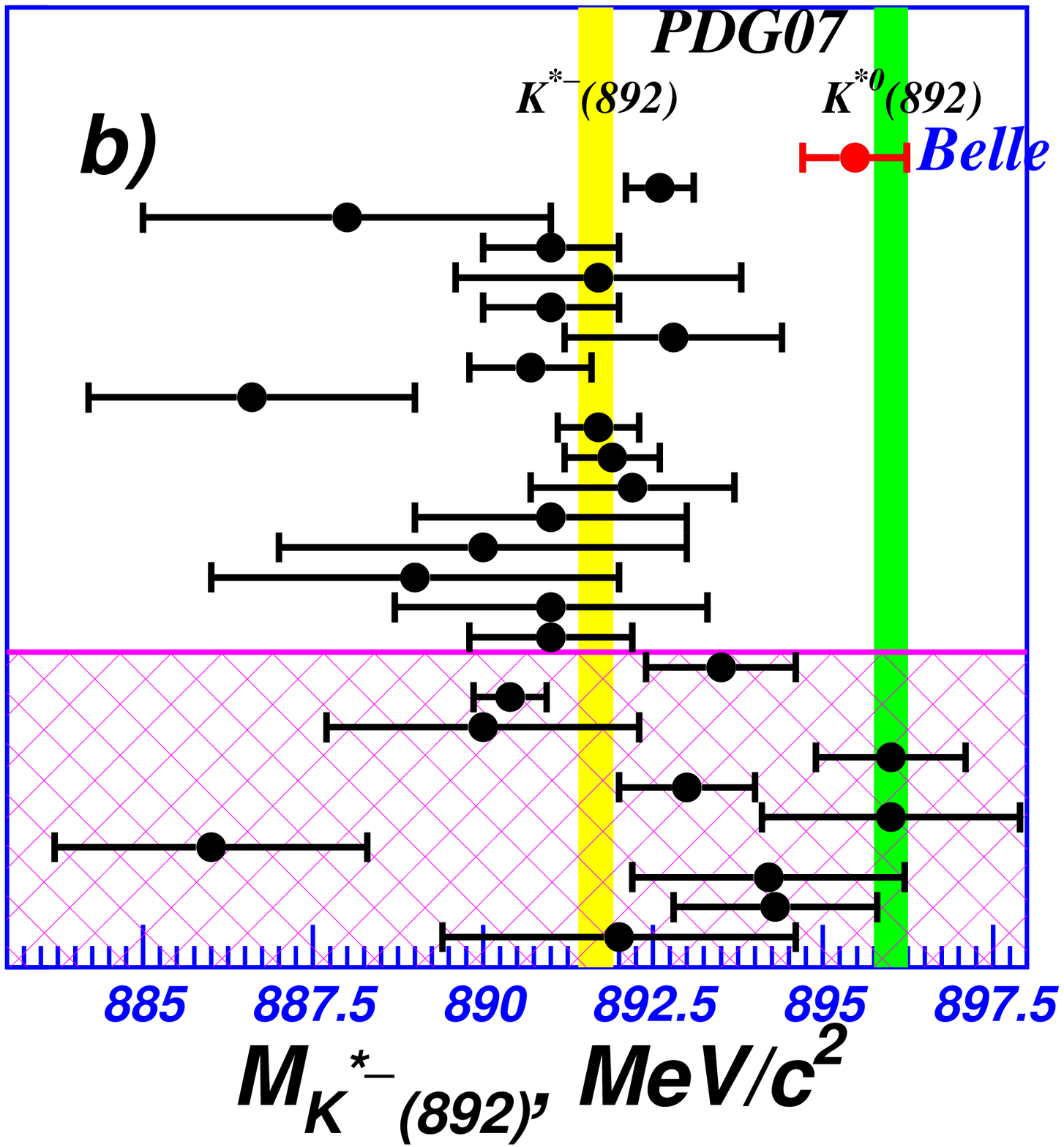}
\\
\parbox[t]{0.95\textwidth}{\caption{ Comparison of the 
$\tau^-\to \bar{K^0}\pi^- \nu_{\tau}$ branching fraction (a) 
and $K^*(892)^-$ mass (b) measured in different experiments.
(b) also shows all available data on the $K^*(892)^-$ mass 
together with the PDG average (the hatched region marks 
PDG data, which were not 
used in the calculation of the average mass, see Ref.~\cite{pdg}), 
as well as our result, which is close to the PDG $K^*(892)^0$ mass.   
\label{compa}}}
\end{figure}

The branching fraction of the $\tau^-\to K_S\pi^- \nu_{\tau}$
decay has been measured using a data sample of $351.4\ {\rm fb^{-1}}$ 
collected with the Belle detector. Our result is:
\begin{center}
$\mathcal{B}(\tau^-\to K_S\pi^-\nu_{\tau})=(0.404\pm 0.002({\rm
stat.})\pm 0.013({\rm syst.}))\%$
\end{center}
{\noindent To compare our result with the previous measurements made  
by the OPAL~\cite{opone}, ALEPH~\cite{alone,althr}, CLEO~\cite{clone} and 
L3~\cite{lthro} groups we calculate the
$\tau^-\to \bar{K^0}\pi^-\nu_{\tau}$ branching 
fraction according to the formula
$\mathcal{B}(\tau^-\to \bar{K^0}\pi^-\nu_{\tau})=\mathcal{B}(\tau^-\to
K_S\pi^-\nu_{\tau})+\mathcal{B}(\tau^-\to
K_L\pi^-\nu_{\tau})=2\mathcal{B}(\tau^-\to K_S\pi^-\nu_{\tau})$ 
and obtain:}
\begin{center}
$\mathcal{B}(\tau^-\to \bar{K^0}\pi^-\nu_{\tau})=(0.808\pm 0.004({\rm
stat.})\pm 0.026({\rm syst.}))\%$
\end{center}
Figure \ref{compa} (a) shows the results of various measurements  
of the $\tau^-\to \bar{K^0}\pi^-\nu_{\tau}$ branching fraction, 
along with the Particle Data Group (PDG) fit value ($\mathcal{B}_{PDG}(\tau^-\to
\bar{K^0}\pi^-\nu_{\tau})=(0.900\pm 0.040)\%$) \cite{pdg} and our result. Our result
is consistent with previous measurements, but is more precise. 

The $K^*(892)$ alone is not sufficient to describe the 
$K_S\pi$ invariant mass spectrum. The best description is achieved in 
the $K_0^*(800)+K^*(892)+K_0^*(1410)$ and $K_0^*(800)+K^*(892)+K_0^*(1430)$ 
models. 
Future high precision studies of the invariant mass
spectra in $\tau$ lepton decays with kaons combined with 
angular analysis, i.e. an application of the structure function
formalism suggested in Ref.~\cite{stone}, will elucidate the nature of
the scalar form factor. They will also check various theoretical models
describing the scalar $K \pi$ sector, e.g., the predictions of the 
resonance chiral theory~\cite{jamin} and the parameters of the
$K^*_0(800)$ resonance calculated from the Roy-Steiner representations
in a model-independent way~\cite{desc}. 

The product of $\tau^-\to K^*(892)^-\nu_{\tau}$ and $K^*(892)^-\to K_S\pi^-$
branching fractions is found to be:
\begin{center}
$\mathcal{B}(\tau^-\to K^*(892)^-\nu_{\tau})\cdot\mathcal{B}(K^*(892)^-\to K_S\pi^-)=(3.77\pm 0.02({\rm stat.})\pm 0.12({\rm syst.})\pm 0.12({\rm mod.}))\times 10^{-3}$,
\end{center}
{\noindent also the $K^*(892)^-$ mass and width are measured:}
\begin{center}
$M(K^*(892)^-)=(895.47\pm 0.20({\rm stat.})\pm 0.44({\rm syst.})\pm
0.59({\rm mod.}))\ {\rm MeV}/c^2$ 

$\Gamma(K^*(892)^-)=(46.2\pm 0.6({\rm stat.})\pm 1.0({\rm syst.})\pm
0.7({\rm mod.}))\ {\rm MeV}$
\end{center}

{\noindent The values of the $K^*(892)^-$ mass and width that we obtain are more
precise than any of the existing measurements of these quantities listed 
in Ref.~\cite{pdg} and shown in Fig.~\ref{compa} (b).} 
While our determination of the width is compatible with most of the previous
measurements within experimental errors, our mass value is 
systematically higher than those before and is in fact consistent with
the world average value of the neutral $K^*(892)^0$ mass, which is
$(896.00 \pm 0.25)$~MeV/$c^2$~\cite{pdg}.
Note that all earlier mass measurements listed in Ref.~\cite{pdg}
come from analysis of hadronic reactions and include 
the effects of final state interaction  while our work presents a 
measurement based on $\tau^-$ decays, where the decay products of the 
$K^*(892)^-$ are the only hadrons involved. It is also noteworthy 
that none of the previous measurements in Ref.~\cite{pdg}, all of which 
were performed more than 20 years ago, present the systematic uncertainties
for their measurements. Unfortunately, previous studies of the
$K^*(892)^-$ in $\tau^-$ lepton decays usually do not determine its
parameters. The only published result we are aware of is that of
ALEPH~\cite{aleph}, which is consistent with ours.  Its accuracy, however, 
is much worse and no systematic errors are presented, which precludes 
any detailed comparisons. 
A similar $K^*(892)^-$ mass shift of $(+4.7\pm 0.9)\ {\rm MeV}/c^2$
was reported by CLEO \cite{bonvi}, but no dedicated study of this effect
was published.
Future dedicated measurements 
of the $K^*(892)^-$ parameters with high precision are necessary 
to clarify this discrepancy and shed light on the long standing
issue of the electromagnetic mass difference between the charged and
neutral $K^*(892)$~\cite{rgg,luc}.     

\section{Acknowledgments}
We are grateful to M.~Jamin for interesting discussions. 
We thank the KEKB group for the excellent operation of the
accelerator, the KEK cryogenics group for the efficient
operation of the solenoid, and the KEK computer group and
the National Institute of Informatics for valuable computing
and Super-SINET network support. We acknowledge support from
the Ministry of Education, Culture, Sports, Science, and
Technology of Japan and the Japan Society for the Promotion
of Science; the Australian Research Council and the
Australian Department of Education, Science and Training;
the National Science Foundation of China and the Knowledge
Innovation Program of the Chinese Academy of Sciences under
contract No.~10575109 and IHEP-U-503; the Department of
Science and Technology of India;
the BK21 program of the Ministry of Education of Korea,
the CHEP SRC program and Basic Research program
(grant No.~R01-2005-000-10089-0) of the Korea Science and
Engineering Foundation, and the Pure Basic Research Group
program of the Korea Research Foundation;
the Polish State Committee for Scientific Research;
the Ministry of Education and Science of the Russian
Federation and the Russian Federal Agency for Atomic Energy;
the Slovenian Research Agency;  the Swiss
National Science Foundation; the National Science Council
and the Ministry of Education of Taiwan; and the U.S.\
Department of Energy.


\begin{thebibliography}{0}

\bibitem{stone} J.~H.~K\"{u}hn and E.~Mirkes, Z.~Phys. C {\bf 56} (1992) 661, 
Erratum-ibid. C {\bf 67} (1995) 364.

\bibitem{sttwo} R.~Decker, E.~Mirkes, R.~Sauer, Z.~W\c{a}s, Z.~Phys. C {\bf 58} 
(1993) 445. 

\bibitem{stthr} M.~Finkemeier and E.~Mirkes, Z.~Phys. C {\bf 69} (1996) 243.  

\bibitem{spe} E.~Gamiz {\it et al.}, Phys. Rev. Lett. {\bf 94} (2005) 011803.

\bibitem{dorfan} J.~Dorfan {\it et al.} (MARK II Collaboration), Phys. Rev. Lett. 
{\bf 46} (1981) 215.

\bibitem{argus} H.~Albrecht {\it et al.} (ARGUS Collaboration), 
Z. Phys. C {\bf 41} (1988) 1.

\bibitem{cltwo} M.~Battle {\it et al.} (CLEO Collaboration), 
Phys. Rev. Lett. {\bf 73} (1994) 1079.

\bibitem{fione} M.~Finkemeier and E.~Mirkes, Z.~Phys. C {\bf 72}
  (1996) 619.

\bibitem{gone} J.J.~Godina Nava and G.~Lopez Castro, Phys. Rev. D {\bf 52}
 (1995) 2850.

\bibitem{alone} R.~Barate {\it et al.} (ALEPH Collaboration),
  Eur. Phys. J. C {\bf 10} (1999) 1.

\bibitem{clone} T.E.~Coan {\it et al.} (CLEO  Collaboration),
  Phys. Rev. D {\bf 53} (1996) 6037. 

\bibitem{kekb} S.~Kurokawa and E.~Kikutani, Nucl. Instr. Meth. A {\bf  499} 
(2003) 1, and other papers included in this Volume.

\bibitem{natk} Z.~Natkaniec {\it et al.} (Belle SVD2 Group),
Nucl. Instr. Meth. A {\bf 560} (2006) 1. 

\bibitem{bel} A.~Abashian {\it et al.} (Belle Collaboration),
  Nucl. Instr. Meth. A {\bf 479} (2002) 117.

\bibitem{eid} K.~Hanagaki {\it et al.}, Nucl. Instr. Meth. A {\bf 485}
 (2002) 490.

\bibitem{muid} A.~Abashian {\it et al.}, Nucl. Instr. Meth. A {\bf  491} 
(2002) 69.

\bibitem{koral} S.~Jadach and Z.~W\c{a}s, Comp. Phys. Commun. {\bf 85}
  (1995) 453.

\bibitem{tola} Z. W\c{a}s, Nucl. Phys. Proc. Suppl. {\bf 98} (2001) 96.

\bibitem{geant} R.~Brun {\it et al.}, GEANT 3.21, CERN Report
  No. DD/EE/84-1 (1984).

\bibitem{pdg} W.-M.~Yao {\it et al.}, J. Phys. G {\bf 33} (2006) 1.

\bibitem{clf} R.A.~Briere {\it et al.} (CLEO  Collaboration), 
Phys. Rev. Lett. {\bf 90} (2003) 181802.

\bibitem{abli} M.~Ablikim {\it et al.} (BES Collaboration), Phys.\
  Lett.\ B {\bf 633} (2006) 681. 

\bibitem{lass} D.~Aston {\it et al.} (LASS Collaboration),
  Nucl. Phys. B {\bf 296} (1988) 493.

\bibitem{baba} B.~Aubert {\it et al.} (BaBar Collaboration), 
Phys. Rev. D {\bf 72} (2005) 072003.

\bibitem{jamin} M.~Jamin, A.~Pich and J.~Portoles, Phys. Lett. 
B {\bf 640} (2006) 176.

\bibitem{desc} S.~Descotes-Genon and B.~Moussallam, Eur. Phys. J.
C {\bf 48} (2006) 553.

\bibitem{opone} G.~Abbiendi {\it et al.} (OPAL  Collaboration), 
Eur. Phys. J. C {\bf 13} (2000) 213.

\bibitem{althr} R.~Barate {\it et al.} (ALEPH Collaboration),
  Eur. Phys. J. {\bf 4} (1998) 29.

\bibitem{lthro} M.~Acciarri {\it et al.} (L3 Collaboration), 
Phys.\ Lett.\ B {\bf 345} (1995) 93.
    
\bibitem{aleph} R.~Barate {\it et al.} (ALEPH Collaboration),
  Eur. Phys. J. C {\bf 11} (1999) 599.

\bibitem{bonvi} G.~Bonvicini {\it et al.} (CLEO  Collaboration), 
Phys. Rev. Lett. {\bf 88} (2002) 111803. 

\bibitem{rgg} A.~De Rujula, H.~Georgi and S.L.~Glashow,
Phys. Rev. D {\bf 12} (1975) 147.

\bibitem{luc} M.~Aguilar-Benitez {\it et al.} (HBC Collaboration), 
Nucl. Phys. B {\bf 141} (1978) 101.

\end{thebibliography}
\end{document}